\documentclass[letterpaper]{article}

\usepackage{prepr}

\title{On the \fwn: Uncovering Sustainability Tool Support for Requirements Engineering}
\author{
\PREPauthor{Marco Stadler}{Johannes Kepler University Linz, LIT Secure and Correct Systems Lab / Institute of Business Informatics -- Software Engineering}{marco.stadler@jku.at}
\PREPauthor{Pascal Taurer}{Johannes Kepler University Linz, Institute of Business Informatics -- Software Engineering}{}
\PREPauthor{Johannes Sametinger}{Johannes Kepler University Linz, LIT Secure and Correct Systems Lab / Institute of Business Informatics -- Software Engineering}{johannes.sametinger@jku.at}
\PREPauthor{Wesley	K.G. Assunção}{North Carolina State University, Department of Computer Science}{wguezas@ncsu.edu}
\PREPauthor{Michael Riegler}{ENGEL Austria GmbH, Information Security}{michael.riegler@engel.at}
\PREPauthor{Michael Vierhauser}{University of Innsbruck, Department of Computer Science}{michael.vierhauser@uibk.ac.at}
\PREPauthor{Iris Groher}{Johannes Kepler University Linz, Institute of Business Informatics -- Software Engineering}{iris.Groher@jku.at}
\vspace{0.6cm}
}
\setvenue{the \vspace{3pt}\textbf{\\\textbf{34th IEEE International Requirements Engineering Conference (RE)}}}
\setvenueone{the \textbf{34th IEEE International Requirements Engineering Conference (RE)}}
\setyear{2026}
\setdoi{TBA}

\usepackage{graphicx}
\usepackage{textcomp}
\usepackage{xcolor}
\usepackage{tcolorbox}

\usepackage{balance}

\usepackage{graphicx}
\usepackage{xspace}
\usepackage[scaled=.92]{helvet}
\usepackage[T1]{fontenc}
\usepackage[scaled=.92]{helvet}
\usepackage[T1]{fontenc}
\usepackage{booktabs} 
\usepackage{url}
\usepackage{multirow}
\usepackage{rotating}
\usepackage{array}
\usepackage{verbatim}
\usepackage{multirow}
\usepackage{setspace}
\usepackage{tabularx}
\usepackage{url}
\usepackage{pifont}
\usepackage{graphicx}
\usepackage{subcaption}
\usepackage[font=footnotesize]{caption}
\usepackage{array,booktabs}
\usepackage{soul}
\soulregister{\ref}{1}
\soulregister{\cite}{1}
\soulregister{\pageref}{1}
\usepackage{vcell}
\usepackage{pifont}
\usepackage{afterpage}
\usepackage{rotating}
\usepackage{wrapfig}
\usepackage{enumitem}
\usepackage{lipsum}
\usepackage{tikz}
\usepackage{listings}
\usepackage{shadowtext}
\usepackage{etoolbox}
\usepackage{harveyballs}
\usepackage{orcidlink}
\usepackage{fontawesome}
\usepackage{circledsteps}
\usepackage{float}
\usepackage[export]{adjustbox}
\usepackage{hyperref}

\newcommand{\fwn}{JI-RADAR\xspace} 
\newcommand{\fwcomp}[1]{\ttformat{#1\xspace}}

\definecolor{dim-ind-color}{HTML}{44AA99}

\definecolor{myblue}{HTML}{007FFF}
\definecolor{myorange}{HTML}{FF8000}
\definecolor{oliveish}{HTML}{ccbb44}
\definecolor{pinkish}{HTML}{ee6677}
\definecolor{greyish}{HTML}{bbbbbb}
\definecolor{blueish}{HTML}{66ccee}
\definecolor{greenish}{HTML}{6ab797}

\newcommand{\mapecharlocal}[1]{\tikz[baseline=-0.5ex]\node[draw, scale=0.8, fill=myblue, text=white, circle, inner sep=0, minimum size=1.25em] {\fwcomp{#1}};}

\newcommand{\ttformat}[1]{{\texttt{\small #1}}}

\definecolor{alizarin}{rgb}{0.82, 0.1, 0.26}

\newcommand{\ucitem}[1]{
\noindent$\bullet$ \textbf{UC-#1 -- }
\ifthenelse{\equal{#1}{1}}{\textbf{Edge Computing:}\xspace}{\textbf{Multi-Modal Auth. System:}\xspace}%
}

\colorlet{punct}{red!60!black}
\definecolor{background}{HTML}{EEEEEE}
\definecolor{delim}{RGB}{20,105,176}
\colorlet{numb}{magenta!60!black}
\lstdefinelanguage{json}{
    basicstyle=\ttfamily\footnotesize,
    numbers=left,
    numberstyle=\scriptsize,
    stepnumber=1,
    numbersep=8pt,
    showstringspaces=false,
    breaklines=true,
    frame=lines,
    literate=
     *{0}{{{\color{numb}0}}}{1}
      {1}{{{\color{numb}1}}}{1}
      {2}{{{\color{numb}2}}}{1}
      {4}{{{\color{numb}4}}}{1}
      {5}{{{\color{numb}5}}}{1}
      {6}{{{\color{numb}6}}}{1}
      {7}{{{\color{numb}7}}}{1}
      {8}{{{\color{numb}8}}}{1}
      {9}{{{\color{numb}9}}}{1}
      {:}{{{\color{punct}{:}}}}{1}
      {,}{{{\color{punct}{,}}}}{1}
      {\{}{{{\color{delim}{\{}}}}{1}
      {\}}{{{\color{delim}{\}}}}}{1}
      {[}{{{\color{delim}{[}}}}{1}
      {true}{{{\color{numb}{true}}}}{1}
      {]}{{{\color{delim}{]}}}}{1},
}

\newcounter{rqcounter}

\makeatletter
\patchcmd{\thebibliography}
  {\section*{\refname}}%
  {\section*{\refname}}
  {}{}
\makeatother

\definecolor{reviewbg}{RGB}{188, 222, 255} 
\sethlcolor{reviewbg}

\newcommand{\hball}[1]{\raisebox{-0.15em}{\csname harveyBall#1\endcsname}}

\DeclareCaptionLabelFormat{purplelabel}{\textcolor{purple}{#1~#2:}}

\newtcbox{\inlinehlbox}{on line,
  colback=reviewbg, colframe=reviewbg,
  boxrule=0pt, boxsep=0pt,
  left=0pt, right=0pt, top=0pt, bottom=0.75pt,
  enhanced, sharp corners}

  \newtcolorbox{highlighted}{enhanced,
  breakable,
  sharp corners,
  colback=reviewbg,
  colframe=reviewbg,
  boxrule=0pt,
  boxsep=0pt,
  left=1pt,
  right=1pt,
  top=0pt,
  bottom=0pt}

\tcbset{
  myrqbox/.style={
    boxrule=1pt,
    arc=6pt,
    left=2mm,
    right=2mm,
    top=1mm,
    bottom=1mm,
    fonttitle=\bfseries,
  }
}

\begin{document}

\maketitle



\section{Introduction}


\textbf{Context:} Software-intensive systems are integral to nearly all facets of modern society~\cite{sustainScrum}. 
Consequently, both their sustainability and their role in facilitating sustainable processes must be established by design~\cite{lagoNewVisionSoftware2026,laguSUSAF2025}.
Software sustainability is defined as ``the preservation of the long-term and beneficial use of software, and its appropriate evolution, in a context that continuously changes''~\cite{lagoNewVisionSoftware2026}. 
Sustainability in Requirements Engineering (RE) is commonly structured along five dimensions: \textit{Environment} referring to the broader ecological effects, \textit{Technical} covering implementation, maintenance, and long-term usability, \textit{Social} describing the integration within communities and impact on society, \textit{Economic}  concerned with the business-related factors, and finally, \textit{Individual} pertaining to an individual's well-being, safety, and privacy~\cite{laguSUSAF2025,Bambazek2023,betzTosemSusaf,lagoNewVisionSoftware2026}. 
One common approach for integrating sustainability into the RE process is the Sustainability Awareness Framework (SuSAF)~\cite{betzTosemSusaf}.
SuSAF aims to raise awareness of the sustainability effects of software systems within their intended context. It provides guidance, instructions, and questions that support stakeholder discussions during interviews and workshops throughout the RE process~\cite{betzTosemSusaf}.


\textbf{RE Problem \& Motivation:} Regulatory initiatives increasingly require (software) organizations to integrate sustainability into their day-to-day business and operational processes.
The United Nations 2030 Agenda formulated 17 Sustainable Development Goals (SDGs)~\cite{unitednationsTransformingOurWorld2015}, while the EU passed the Corporate Sustainability Reporting Directive (CSRD), which requires companies to publish and audit sustainability-related information~\cite{DirectiveEU20222022}.
Regulations and laws require organizations in the software development sector to disclose both qualitative and quantitative sustainability metrics, among other obligations~\cite{sustainScrum}.
Consequently, integrating sustainability reporting processes into the software development life cycle becomes increasingly important. 
RE processes often lack systematic methods to elicit, analyze, and prioritize sustainability requirements alongside functional and non-functional requirements, and studies indicate that tool support for this integration remains limited~\cite{Bambazek2023}. 

To address this gap, we introduce \fwn, which supports \textbf{stakeholders involved in system design} (e.g., developers, requirements engineers, project managers, and usability engineers)~\cite{betzTosemSusaf}  by providing practical tools to integrate sustainability into the RE process. 
We extend the widely used Atlassian Jira platform~\cite{JiraProjectManagement} by implementing a ready-to-use plugin that can be directly adopted in industrial practice.





\section{The \fwn Tool}



\textbf{Methodology:} The \underline{JIRA} \underline{D}ashboard for \underline{A}ssessing Sustainability \underline{R}equirements (\fwn) is built on the widely adopted project management and issue-tracking tool  ``Jira''~\cite{JiraProjectManagement} and is designed to systematically capture, assess, and trace sustainability-related impacts.
It leverages Jira’s agile development features, in particular Scrum capabilities (backlog, sprint planning, etc.), and extends default components (User Stories, Issues, etc.) with functionality to support sustainability considerations at both the process and issue level.
To align with Jira’s agile aspects, we reuse the SustainScrum extensions introduced in our previous works~\cite{sustainScrum}.
In this work, we integrated sustainability considerations into the agile development process through additional artifacts and activities using Software Engineering Methods and Theory (SEMAT kernel)~\cite{jacobson2012essence}. 
A central element of SustainScrum is the Sustainability Management Model (SuMM), which serves as the primary source for sustainability-related decision-making.
SuMM represents a weighted decision matrix in which the previously defined sustainability dimensions from SusAF provide the context for stakeholders’ priorities.


\textbf{Key Elements:} \citefig{elems} provides an overview of the key artifacts from SustainScrum used in \fwn.
The \emph{SuMM} provides the relevant dimensions and weight configurations from the stakeholders, which form the basis for subsequent sustainability assessments and evaluations.

\begin{figure}
    \centering
    \includegraphics[width=\linewidth]{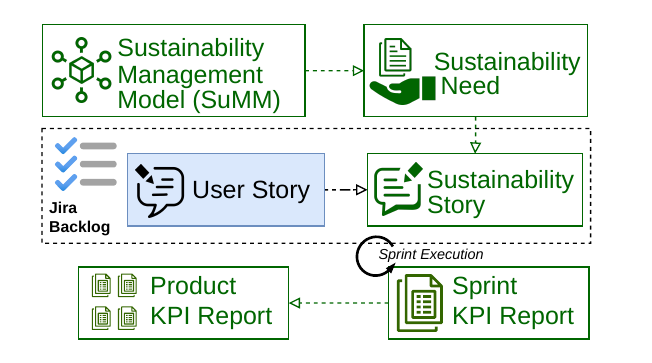}
    \caption{Key Elements of SustainScrum incorporated in \fwn.}
    \label{fig:elems}
\end{figure}

Based on this, \textit{Sustainability Needs} can be derived and presented in an easy-to-read, understandable format, i.e., \textit{Sustainability Stories}, using the same format as User Stories, however, specifically targeting sustainability-related stakeholder needs, serving as traceable justification and requirements artifacts. 
In addition, mappings between \textit{Sustainability Stories} and the \textit{SuMM} dimensions can be represented, and stakeholder decisions (pertaining to sustainability-relevant functionality) are documented, ensuring that negotiated priorities remain transparent and traceable. 
Finally, \emph{Sprint KPI Reports} and aggregated \emph{Product KPI Reports} can be created. The Sprint KPI Report provides an indicator matrix for the sprint scope, while the Product KPI Report aggregates indicators across a release scope. KPI reports aggregate the evaluated stories at the sprint and release/product levels. 

\textbf{Implementation:} \fwn is implemented as a Jira plugin and can be installed on any Jira instance directly via a single-click console-link installation (cf.~\href{https://github.com/jku-win-se/JI-RADAR}{GitHub}). 
The high-level architecture of \fwn is shown in \citefig{architecture}. 
\fwcomp{Stakeholders}~\mapecharlocal{1} interact with the standard \fwcomp{Jira User Interface}~\mapecharlocal{2} to set up the SuMM, define the Sustainability Stories, and analyze generated reports.
These interactions trigger serverless functions in the \fwcomp{Forge App Runtime}~\mapecharlocal{3} (\fwcomp{Resolvers}) via the \fwcomp{Atlassian Cloud API}~\mapecharlocal{4}. 
The SuMM and other plugin-related data are stored on a \fwcomp{Storage} component on the \fwcomp{Atlassian Cloud}.

\textbf{Proof-of-Concept Evaluation and Results:}
As initial evidence of \fwn\!\!\!’s usability, we conducted a pilot user study ($N=4$) in the form of semi-structured interviews. Participants were asked to test and execute certain tasks related to the developed tool and, afterward, to comment on the tool's usability and the systematic support for sustainability assessment. 
Participants reported that the tool significantly improved both the capture and assessment of sustainability information, with both aspects scoring a median of 4.5 of 5.

\section{Related Work}
\label{sec:relwork}

The need for technical solutions to support sustainability frameworks, like SuSAF, has led to several prototype tools. SuSoftPro~\cite{Alharthi2018} is a web-based tool for requirements engineers that uses Fuzzy Rating Scales and the TOPSIS method to provide a quantitative analysis of requirements' impacts. 
SusApp~\cite{basmer2021}, designed specifically for SusAF, simplifies the documentation and visualization of sustainability effects and automates the generation of SusAD diagrams. 
Scrum Sustainability Poker~\cite{Bambazek2024}, inspired by the traditional planning poker method, facilitates team discussions during backlog refinement to estimate the potential sustainability impacts of individual user stories.
Despite these developments, practitioners have identified a significant gap in tool integration~\cite{Bambazek2023}. Managing sustainability data in standalone applications alongside primary project management tools is often perceived as cumbersome. 
\fwn contributes to this gap by integrating sustainability into Atlassian Jira. 

\begin{figure}[t!]
    \centering
    \includegraphics[width=\linewidth]{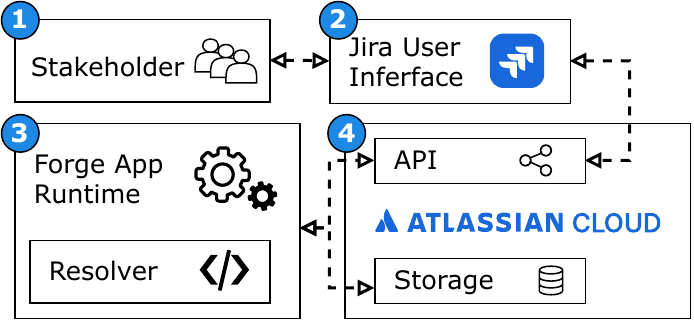}
    \caption{High-level overview of \fwn's architecture.}
    \label{fig:architecture}
\end{figure}

\section{Conclusion}
\label{sec:conclusion}

In this paper, we present \fwn, a tool for integrating sustainability metrics directly into (agile) RE. It demonstrates that tool support can effectively bridge the gap between abstract regulatory reporting requirements (such as the CSRD) and day-to-day development workflows without forcing stakeholders to abandon established platforms like Jira.



\bibliographystyle{abbrv}
\bibliography{main}

\newpage
\section*{Annex}



In the Annex, we further outline how we intend to present \fwn at the RE conference. 

\textbf{Data availability:} \fwn's code, together with further instructions for installation and for building the plugin from source, is publicly available on GitHub.\footnote{\url{https://github.com/jku-win-se/JI-RADAR}}

\textbf{Video showcase:} We uploaded a video showcasing the steps outlined below to YouTube.\footnote{\url{https://youtu.be/MPToigRrNzU}}

For the live demo, we plan to hand out a laptop to interested parties and guide them through the four steps while they experience the tool hands-on. 
We structure the presentation of \fwn into three steps:
\begin{enumerate}[leftmargin=3.5em]

\item[\textit{Step 0:}]\textit{Installation} -- In the first part, we will briefly explain and demonstrate what the stakeholders and users are required to do in order to install \fwn before it can be used in a Jira instance.

\item[\textit{Step 1:}] \textit{Setup} -- In this part, we will showcase how \fwn can be configured and tailored to a given use case and project setup.

\item[\textit{Step 2:}] \textit{Assessment of Jira Artifacts} -- In the second part, we will demonstrate the core features of \fwn. This includes a detailed description of how the tool can be used during the day-to-day business of RE and how and where it extends Jira.

\item[\textit{Step 3:}] \textit{Reporting} -- In the last part, we explain which insights and reports \fwn can produce to analyze, synthesize, and plan next steps.
\end{enumerate}

\subsection{Step 0: Installation}
Since \fwn is a Jira plugin, a Jira instance is required to install it.
For this, an Atlassian account must be created. Afterward, a Jira workspace can be created. \newline
\emph{For the live presentation, we will set up multiple Jira (Atlassian) accounts in the Atlassian cloud to showcase how easy it is to get \fwn up and running.}\newline
Once the user is logged into an Atlassian account and has created a Jira workspace, \fwn can be installed in the Jira instance.

The \fwn plugin is available for installation online (cf.~link on  \href{https://github.com/jku-win-se/JI-RADAR/tree/main}{GitHub}); We followed the instructions on Atlassian's app distribution page.\footnote{\url{https://developer.atlassian.com/platform/forge/distribute-your-apps}}
The link leads to the installation page depicted in \citefig{install}.
After reviewing the permissions, the user simply needs to click the ``Get app'' button and confirm the installation workspace (i.e., select the desired Jira workspace). 
This will trigger the installation, and after a few seconds, \fwn will be ready to use.

\begin{figure}[!ht]
    \centering
    \includegraphics[width=\linewidth,frame]{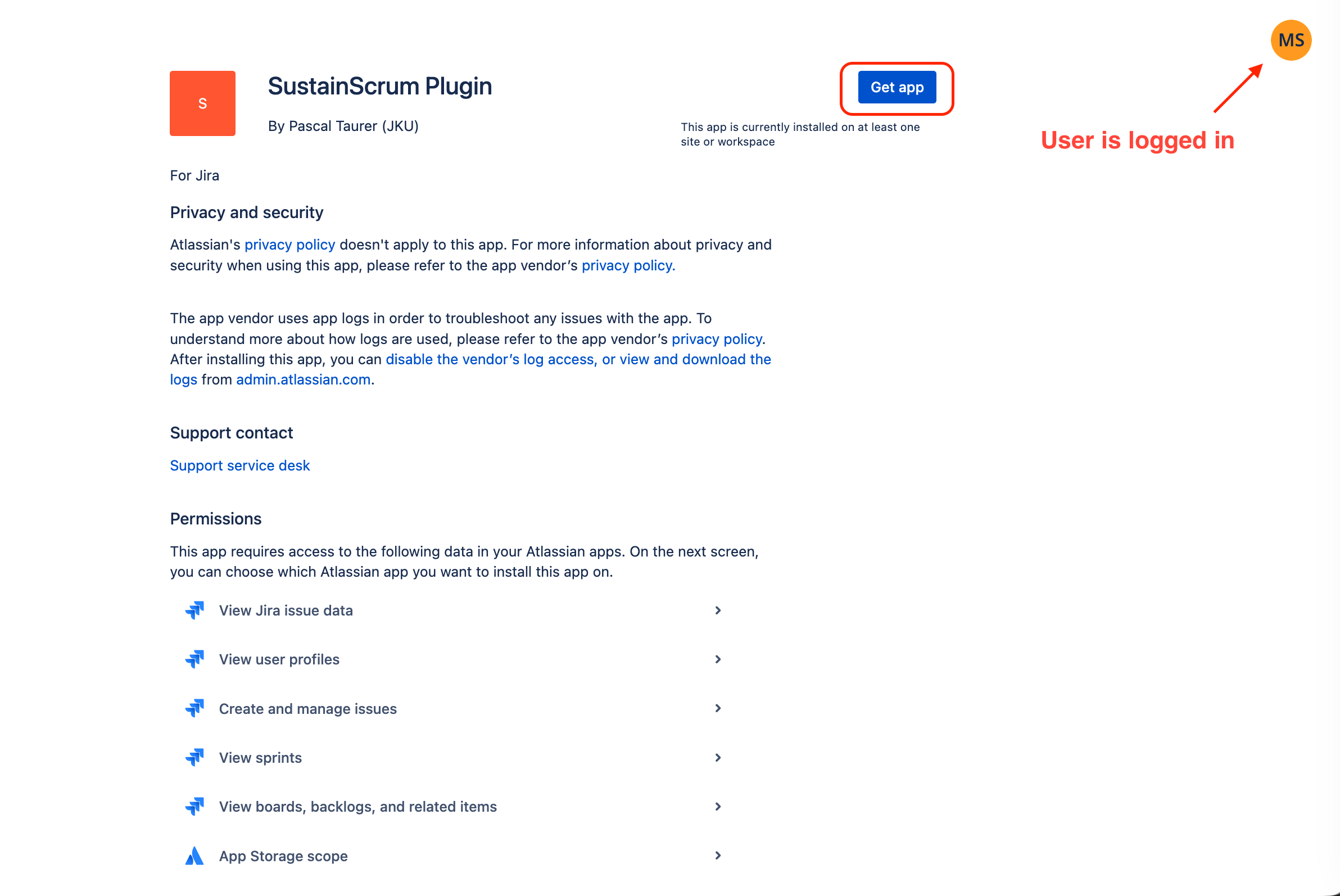}
    \caption{Installation page for our plugin with the one-click installation button.}
    \label{fig:install}
\end{figure}

\subsection{Step 1: Setup}
Once the \fwn plugin is installed, the sustainability aspects need to be configured before it can be used for a specific project and during sprints.
Not every sustainability dimension is equally important for every project, stakeholder, or use case. 
\fwn thus allows adaptation of the dimensions and their importance to specific stakeholder needs.
This is done using SuMM dimensions and weights. 

The \fwcomp{SuMM Configuration} (located in the drop-down menu) is used to customize \fwn and to define project-specific sustainability focus points.  In the \fwcomp{SuMM Configuration} menu, we can select which dimensions to include or exclude when evaluating and documenting the sustainability impacts of Jira artifacts. 
It is also possible to assign weights to the dimensions.
This configuration step is depicted in \citefig{summ-dim}.
For instance, if privacy plays a more prominent role in a financial software application, it makes sense to increase the weight of the ``individual'' domain to emphasize this factor in the calculation of the sustainability impacts and KPIs.

\begin{figure}[b!]
    \centering
    \includegraphics[width=\linewidth,frame]{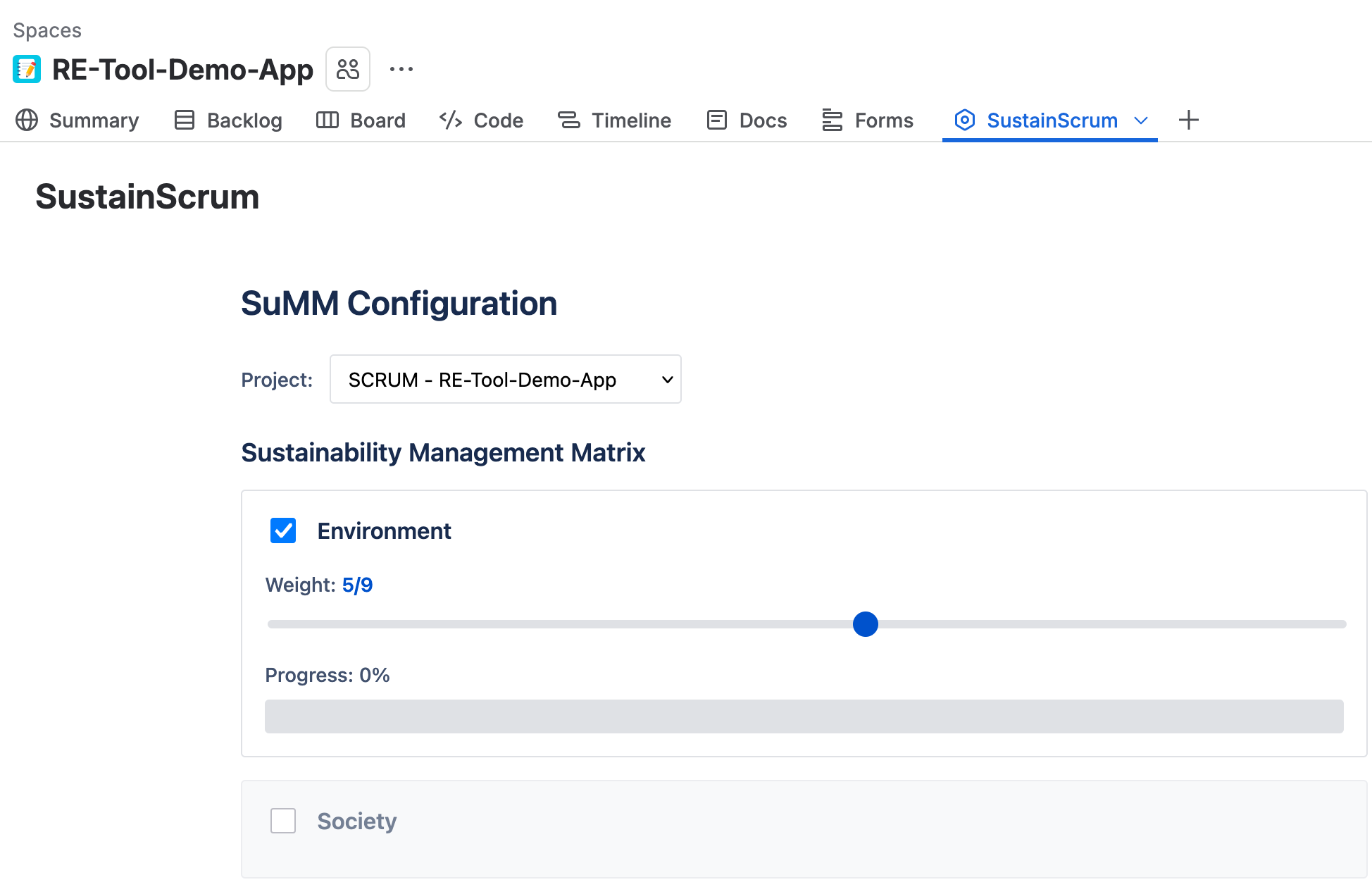}
    \caption{Selection of relevant dimensions and weight assignments for the \fwcomp{SuMM Configuration}.}
    \label{fig:summ-dim}
\end{figure}

For every artifact involved in a sustainability assessment, the plugin asks a series of questions that need to be answered using an adjusted Likert scale (cf. details in \textit{Step~2}).
The default questions configured and used in \fwn are based on and inspired by SuSAF. 
However, \fwn allows us to add, remove, and modify questions for each dimension.
This makes \fwn highly configurable, allowing users to add domain- and project-specific sustainability questions. 
\citefig{summ-questions} depicts a screenshot with the configurable questions for the ``Environment'' and ``Society'' dimensions.

\begin{figure}[t!]
    \centering
    \includegraphics[width=\linewidth,frame]{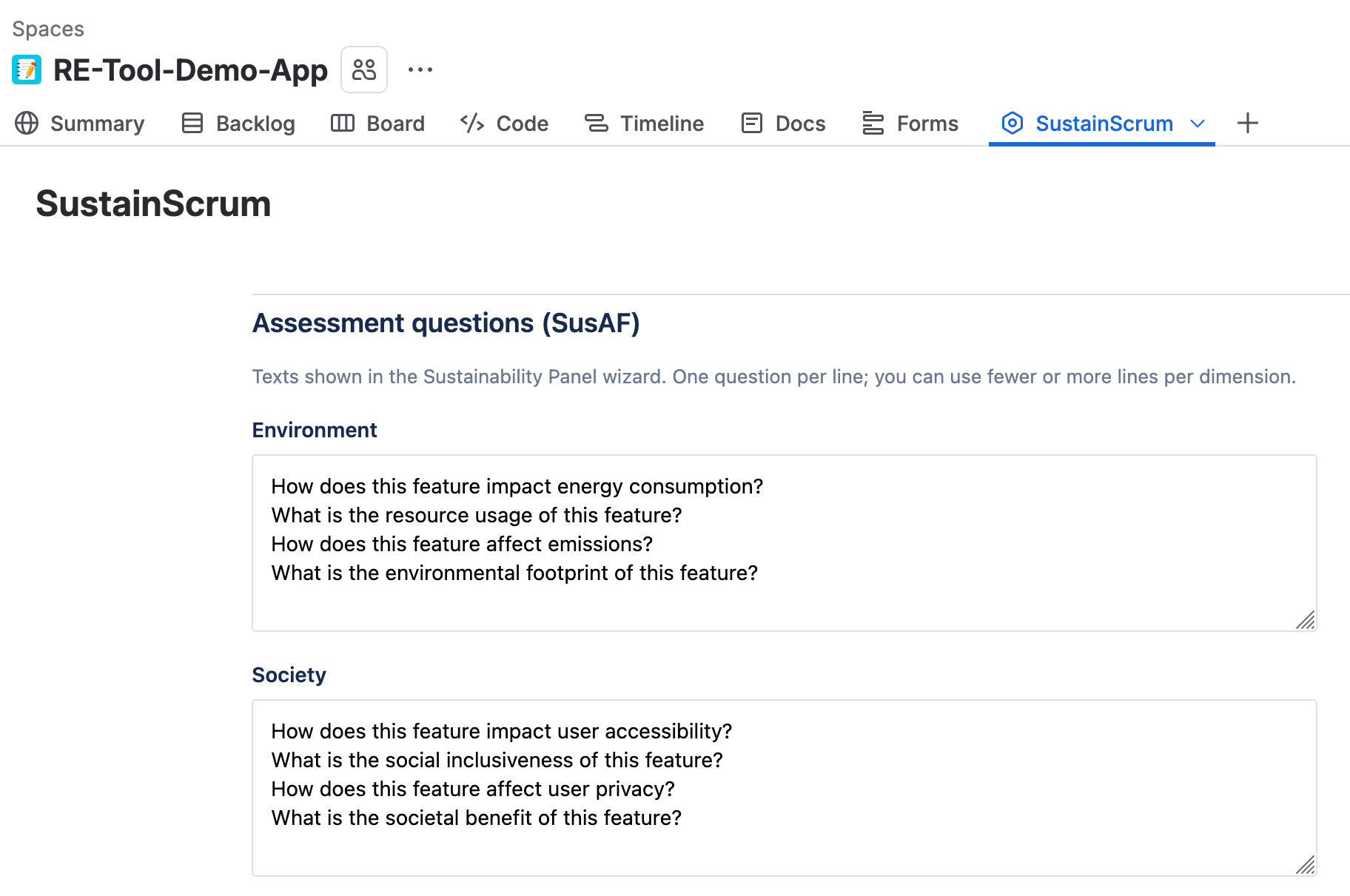}
    \caption{Customization of the assessment questions in the SuMM.}
    \label{fig:summ-questions}
\end{figure}

Although configuring the SuMM for a specific use case, incorporating stakeholder sustainability needs, it is also possible to skip the personalization setup and use the plugin right away with the default configuration. Once the SuMM configuration is complete, clicking on the ``Save Configuration'' button completes this step.

\subsection{Step 2: Assessment of Jira Artifacts}
\label{sec:p2}

After configuring the SuMM, the plugin is ready  for use in  the sustainability assessment. For every artifact in Jira (e.g., Tasks, User Stories, Epics), a new section, \fwcomp{Sustainability Assessment} (cf.~\citefig{assess-sec}), is added.

\begin{figure}[b!]
    \centering
    \includegraphics[width=\linewidth,frame]{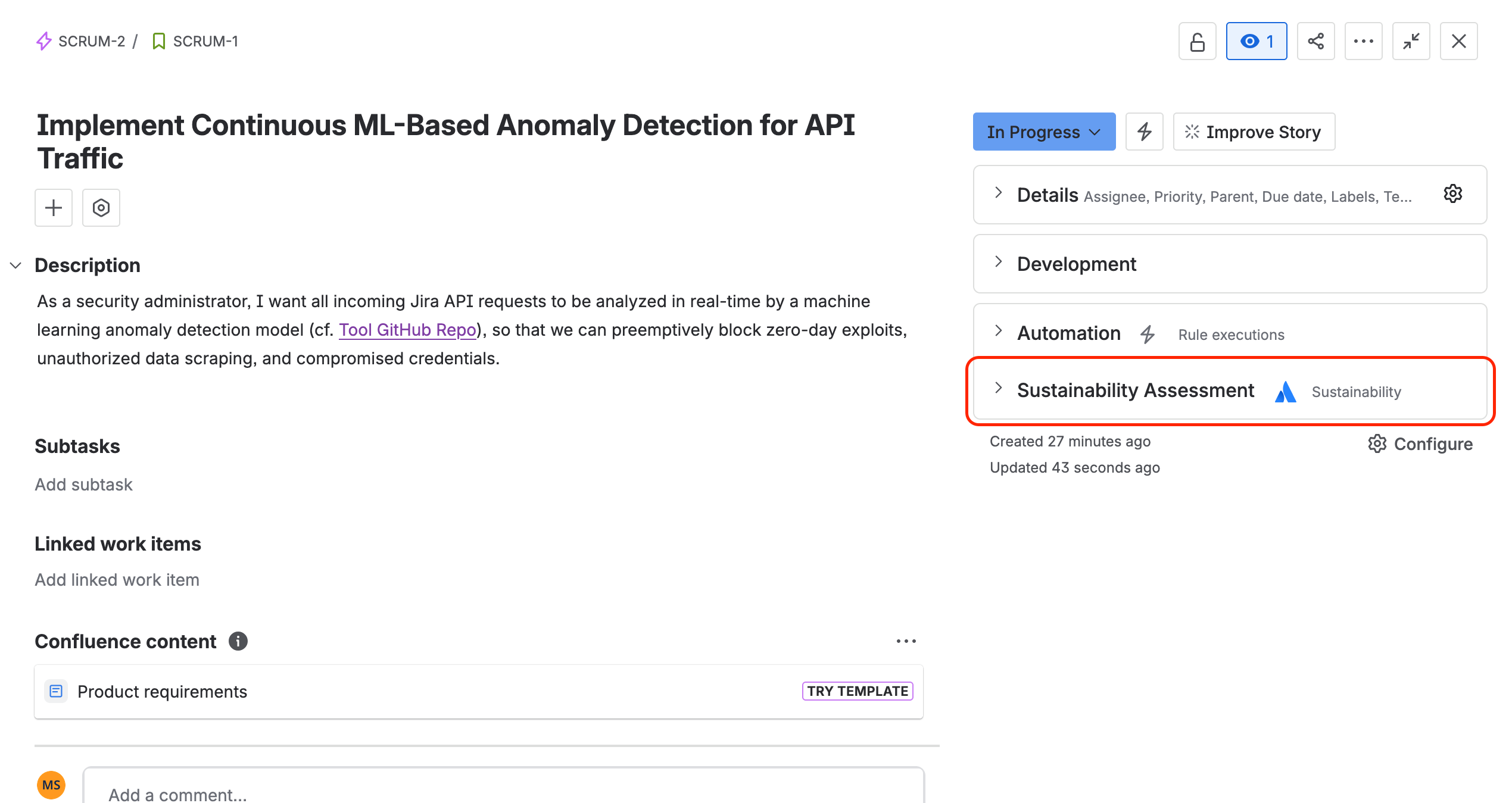}
    \caption{Assessment section for a user story.}
    \label{fig:assess-sec}
\end{figure}

The subsequent \fwcomp{Sustainability Assessment} is a ``wizard'',  that guides  users through the questions specified in the SuMM.
For each selected dimension, and question part of the dimension, the user has to assess the impact of the Jira artifact (cf.~\citefig{ass-quest-wiz}).
They can select a value from an adjusted Likert scale (1=Direct Negative Impact, 2=Indirect Negative Impact, 3=No Impact, 4=Indirect Positive Impact, 5=Direct Positive Impact, Indifferent=Not applicable / Cannot assess).

After the (mandatory) assessment of the questions for each dimension is conducted, users can (optionally) document trade-offs, alternatives, and rationales for the given sustainability assessment. Furthermore, users can link related issues (e.g., conflicting user stories) to establish traceability.

\begin{figure}[t!]
    \centering
    \includegraphics[width=\linewidth,frame]{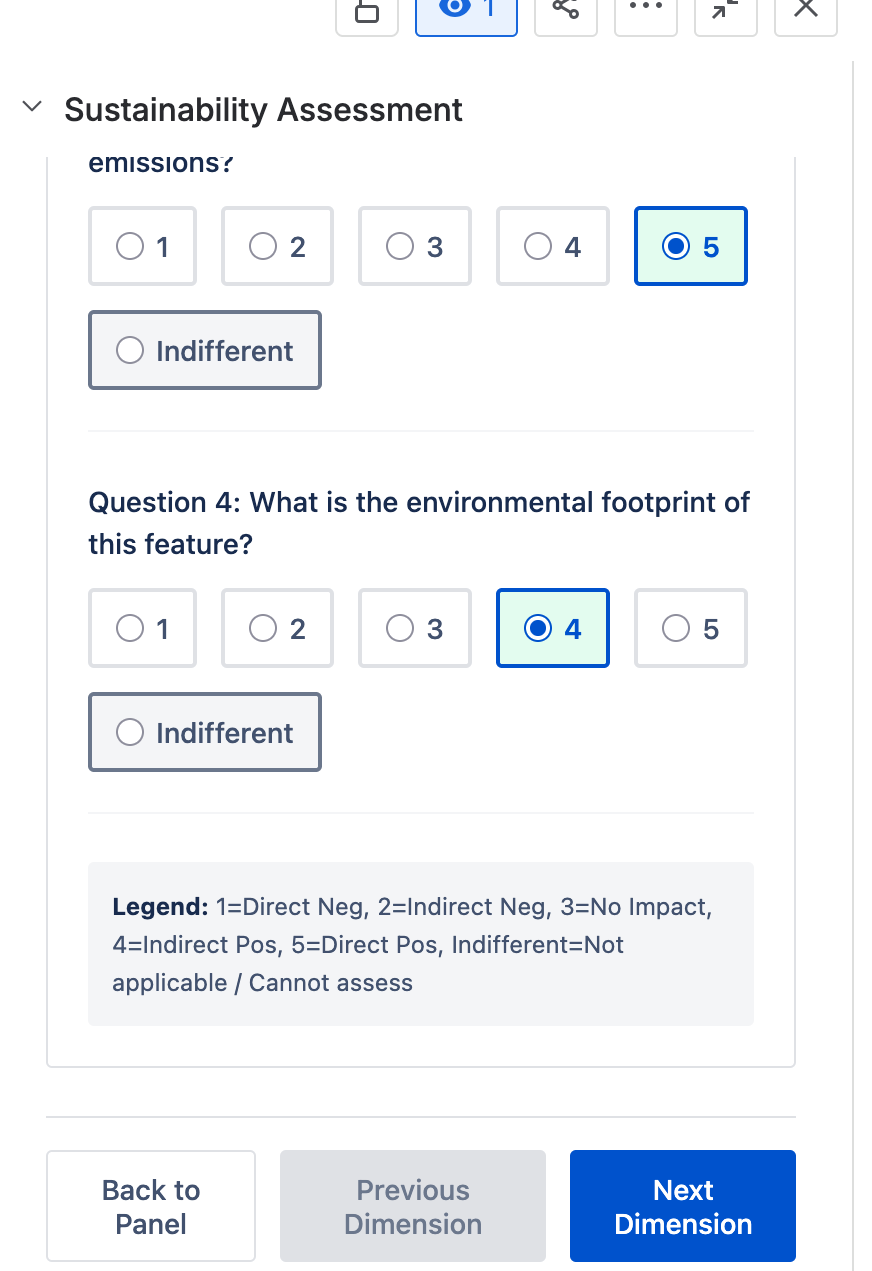}
    \caption{Assessment wizard with example to be answered by the user.}
    \label{fig:ass-quest-wiz}
\end{figure}

\subsection{Step 3: Reporting}
After assessing the artifacts, the (aggregated) results of the sustainability assessment can be viewed on the \fwcomp{Sustainability Dashboard} page.
The dashboard provides a series of views, filters, and insights.
For example, the dashboard provides (weighted) KPIs, heatmaps, trend graphs, various filter options (e.g., by sprint or dimension), and the option to export the results (e.g., into a CSV file).
All these features contribute to the goal of providing a holistic overview of the sustainability assessment  and the Jira workspace's impact.
\citefig{dashboard} depicts the dashboard's overview tab.

\begin{figure}[b!]
    \centering
    \includegraphics[width=\linewidth,frame]{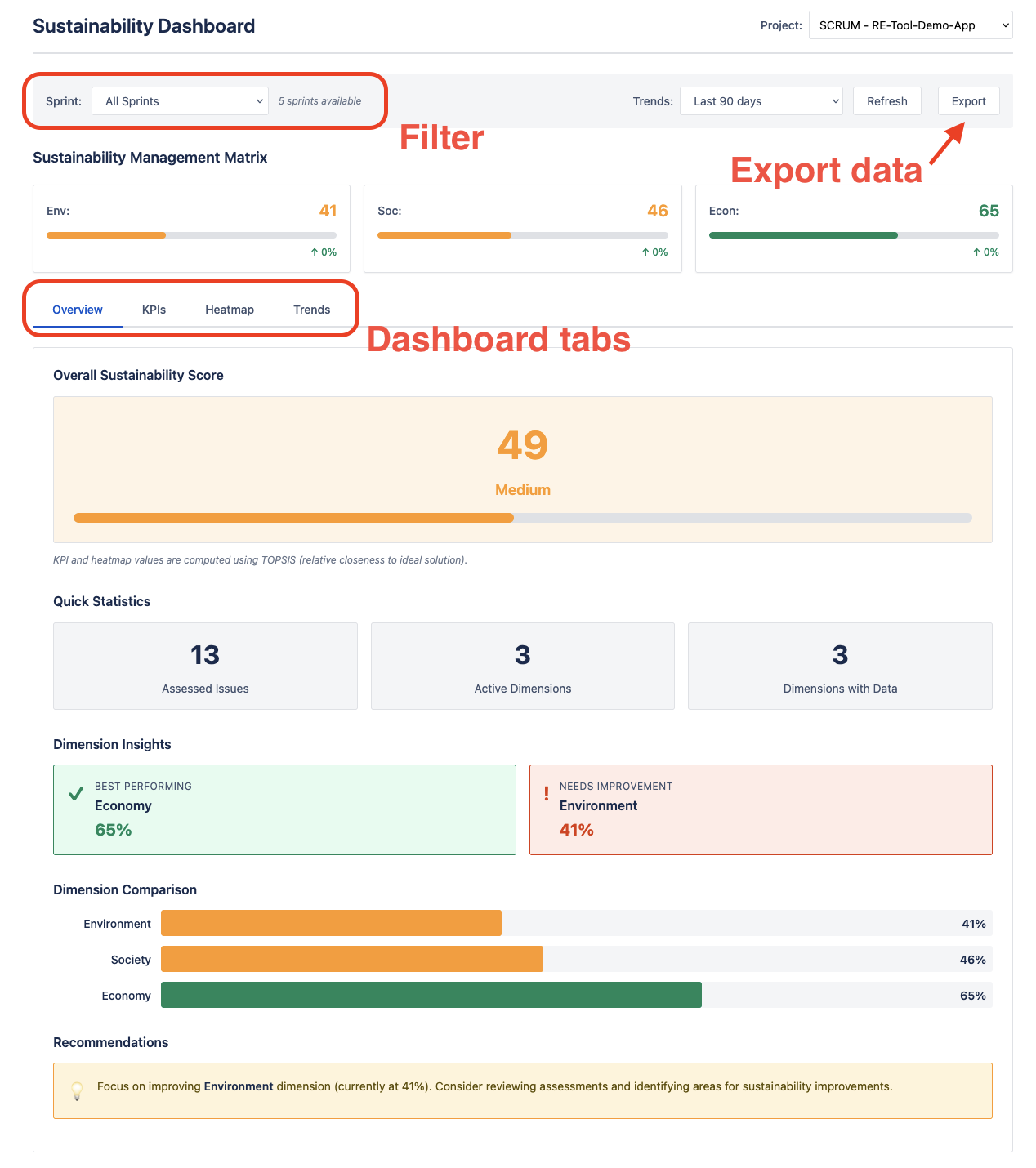}
    \caption{Dashboard overview with different filter and export options.}
    \label{fig:dashboard}
\end{figure}



\end{document}